\documentclass[envcountsame]{llncs}
\usepackage{cleveref}
\usepackage{graphicx}
\usepackage{semantic}
\usepackage{tikz}
\usepackage{xcolor}
\usepackage{mdframed}
\usepackage{amssymb}
\usepackage{pifont}
\usepackage{multirow}
\usepackage{wrapfig}
\usepackage{listings}

\newcommand*{\ttfamilywithbold}{\fontfamily{lmtt}\selectfont}

\usetikzlibrary{shapes,arrows,positioning,decorations.markings,matrix,fit,backgrounds}
\tikzstyle{line} = [draw, -latex']
\pgfdeclarelayer{background}
\pgfdeclarelayer{foreground}
\pgfsetlayers{background,main,foreground}

\newsavebox{\dratlfscbox}
\newsavebox{\erlfscbox}

\usepackage{xspace}
\newcommand{\set}[1]{\left\{#1\right\}}

\newcommand{\tup}[1]{\left<#1\right>}
\newcommand{\til}{,\ldots,}

\newcommand{\minisat}{MiniSat\xspace}
\newcommand{\cryptominisat}{Crypto\-Mini\-Sat\xspace}
\newcommand{\drat}{DRAT\xspace}
\newcommand{\drattrim}{DRAT-trim\xspace}
\newcommand{\drattoer}{drat2er\xspace}
\newcommand{\lrat}{LRAT\xspace}
\newcommand{\lfsc}{LFSC\xspace}
\newcommand{\er}{ER\xspace}
\newcommand{\lf}{LF\xspace}
\newcommand{\cvcfour}{CVC4\xspace}
\newcommand{\coq}{Coq\xspace}
\newcommand{\smtcoq}{SMTCoq\xspace}

\newcommand{\bi}{\begin{itemize}}
\newcommand{\ei}{\end{itemize}}
\newcommand{\be}{\begin{enumerate}}
\newcommand{\ee}{\end{enumerate}}
\newcommand{\bd}{\begin{description}}
\newcommand{\ed}{\end{description}}

\newcommand{\liff}{\leftrightarrow}
\newcommand{\op}{\ensuremath{\diamond}\xspace}

\begin{document}
%
\title{DRAT-based Bit-Vector Proofs in CVC4\thanks{This work was supported in part by DARPA (N66001-18-C-4012 and FA8650-18-2-7861) and NSF (1814369).}}

\author{Alex Ozdemir 
	\and
Aina Niemetz
	\and
Mathias Preiner 
\and
Yoni Zohar  
\and
Clark Barrett
\institute{Stanford University 
}\\
}
\maketitle              
\begin{abstract}
  Many state-of-the-art Satisfiability Modulo Theories (SMT) solvers for
  the theory of fixed-size bit-vectors employ an approach called
  bit-blasting, where a given formula is translated into a Boolean
  satisfiability (SAT) problem and delegated to a SAT solver.
  Consequently,
  producing bit-vector proofs in an SMT solver requires incorporating
  SAT proofs into its proof infrastructure.
  %
  In this paper,
  we describe three approaches for integrating DRAT proofs generated by an
  off-the-shelf SAT solver into the proof infrastructure of the SMT solver CVC4
  and explore their strengths and weaknesses.
  We implemented all three approaches using
  \cryptominisat as the SAT back-end for its bit-blasting engine
  and evaluated performance in terms of proof-production and
  proof-checking.
\end{abstract}
\section{Introduction}

The majority of Satisfiability Modulo Theories (SMT) solvers for the
theory of fixed-size bit-vectors employ an approach called
bit-blasting. That is, an input formula is first simplified, and then eagerly
translated into propositional logic and handed to a Boolean satisfiability
(SAT) solver.
Thus,
when producing a proof of unsatisfiability for a given bit-vector input, it is
crucial to obtain the unsatisfiability proof from
the SAT solver back-end and incorporate it into a possibly larger SMT proof.
The bit-blasting engine of the
SMT solver \cvcfour~\cite{CVC4}
currently supports several SAT solvers as back-ends.
Producing proofs, however, is only supported with a modified version of
\minisat~\cite{minisat},
which was extended to record resolution proofs that can be embedded into
\cvcfour proofs~\cite{DBLP:conf/lpar/HadareanBRTD15}.
This custom \minisat
implementation requires extra maintenance and is less competitive
than more recent off-the-shelf SAT solvers.

In recent years, the \emph{Delete Resolution Asymmetric Tautologies} (\drat)
proof system~\cite{drattrim}, a generalization of
\emph{extended resolution} (ER)~\cite{Tseitin1983},
has become the de facto standard for validating unsatisfiability in SAT solvers.
Using a state-of-the-art SAT solver with support for DRAT inside \cvcfour
would allow \cvcfour to use the latest, best SAT techniques
while being able to produce bit-vector proofs without additional
customization of the SAT solver code.
However, in order to support this, \cvcfour must
be able to incorporate \drat proofs into its proof infrastructure,
which is based on \lfsc, an extension of
\emph{Edinburgh's Logical Framework}~\cite{Harper1993} (LF)
with functional programs called \emph{side conditions} (see~\cite{Stump2013} for more details on \lfsc and \cite{BdMF15} for a more general survey of proofs in SMT-solvers).
In this paper,
we examine three approaches for translating \drat proofs to
\lfsc:
$(i)$ a direct translation from \drat to \lfsc proofs,
$(ii)$ an intermediate translation from \drat to Linear RAT (\lrat)
    proofs~\cite{cruz2017efficient},
    and
$(iii)$ an intermediate translation from \drat to \er
    proofs~\cite{kiesl2018extended},
    which are then translated to \lfsc.
The produced proofs can be independently checked
by any proof checker for \lfsc.
We describe the implementation of these three approaches for
generating bit-vector proofs in \cvcfour, 
discuss their strengths and weaknesses,
and evaluate their performance in terms of proof production and
proof checking.

\section{From \drat to \lfsc}

We briefly review the definitions relevant to the proof systems \drat, \lrat, and \er.
More details can be found in \cite{cruz2017efficient,kiesl2018extended,drattrim}.

A \emph{literal} is either a propositional variable or its negation.
A \emph{clause} is a disjunction of literals, sometimes interpreted as a set of
literals.
A clause is \emph{unit} if it is a singleton.
A \emph{formula} in conjunctive normal form (CNF) is a conjunction of clauses,
sometimes interpreted as a set of clauses.

A proof for formula $F$ in CNF is a
sequence $\pi=C_{1}\til C_{m},\allowbreak I_{m+1}\til I_{n}$
with clauses
$C_{1}\til C_{m}\in F$ and
pairs $I_{i}$ of the form $\tup{\op,X}$,
where
either $\op\in\set{a,d}$ and $X$ is a clause, or $\op=e$ and $X$ is a CNF
formula.
Letters $a$, $d$, and $e$ indicate addition, deletion, and extension, respectively.
Sequence $\pi$ induces a sequence of CNF formulas
$F_{0}\til F_{n}$ such that $F_{i}=\set{C_{1}\til C_{i}}$ for $1\leq i\leq m$,
and for $i>m$,
$F_{i}=F_{i-1}\cup\set{C}$ if $I_{i}=\tup{a,C}$,
$F_{i}=F_{i-1}\setminus\set{C}$ if $I_{i}=\tup{d,C}$, and
$F_{i}=F_{i-1}\cup G$ if $I_{i}=\tup{e,G}$.
It is a proof of unsatisfiability of $F$ if $\emptyset \in F_n$.

A proof $\pi$ of unsatisfiability of $F$ is a valid \er proof if every
$I_{i}$ is either: (i)
$\tup{a,C\cup D}$,
where $C\cup\set{p}$ and $D\cup\set{\overline{p}}\in F_{i-1}$ for some $p$;
or (ii) $\tup{e,G}$,
where $G$ is the CNF translation of $x \liff \varphi$ with $x$ a
fresh variable and $\varphi$ some formula over variables occurring in $F_{i-1}$.
Proof $\pi$ is a valid \drat proof if every $I_{i}$ is
either $\tup{d,C}$ or
$\tup{a,C}$ and for the latter, one of the following holds:
\vspace*{-1ex}
\bi
\item $C$ is a \emph{reverse unit propagation}
  (RUP)~\cite{DBLP:conf/isaim/Gelder08} in $F_{i-1}$, i.e., the empty clause
  is derivable from $F_{i-1}$ and the negations of literals in $C$ using unit
  propagation.
\item $C$ is a \emph{resolution asymmetric tautology} (RAT)
  in $F_{i-1}$, i.e.,
  there is some $p\in C$ such that for every $D\cup\set{\overline{p}}\in
  F_{i-1}$, $C\cup D$ is a RUP in $F_{i-1}$. If $C$ is a RAT but not a RUP, we call it a
  \emph{proper} RAT.
\ei
\vspace*{-1ex}

\noindent
\lrat proofs are obtained from \drat proofs by allowing a third element in each $I_{i}$ 
that includes hints regarding the clauses and literals that are relevant 
for verifying the corresponding proof step.

\subsection{Integration Methods}
\label{sec:integration}

The \textit{Logical Framework with Side Conditions} (\lfsc) \cite{Stump2013} is a
statically and dependently typed Lisp-style meta language
based on the Edinburgh Logical Framework (\lf)~\cite{Harper1993}.
It can be used to define logical systems and check proofs written
within them
by way of the Curry-Howard correspondence.
Like
\lf, \lfsc is a framework in which axioms and derivation rules can be defined
for multiple theories and their combination.
\lfsc additionally adds the notion of \textit{side conditions} as
functional programs, which can restrict the application of derivation rules.
This is convenient for expressing proof-checking rules
that are computational in nature.
In order to use \drat proofs in \cvcfour,
the proofs need to be representable in \lfsc.
We consider the following three approaches for integrating \drat proofs into \lfsc.

\vspace*{1ex}\noindent
\textbf{Checking \drat Proofs in \lfsc.}
This approach directly translates \drat proofs into \lfsc.
It
requires creating a signature for \drat in \lfsc, which essentially is an
\lfsc implementation of a \drat checker.


\vspace*{1ex}\noindent
\textbf{Checking \lrat Proofs in \lfsc.}
\lrat proofs include hints to accelerate unit propagation while proof
checking.  We use the tool \drattrim~\cite{drattrim} to translate \drat proofs
into the \lrat format and then check the resulting proof with an \lrat \lfsc
signature.

\vspace*{1ex}\noindent
\textbf{Checking \er Proofs in \lfsc.}
This approach aims at further reducing computational overhead
during proof checking by translating a \drat proof into an \er proof with the
tool \drattoer~\cite{kiesl2018extended}.
The \er proof is then translated to \lfsc and checked with an \er
\lfsc signature.



\section{\lfsc Signatures}
\label{sec:sigs}

\begin{figure}[t]
\centering
  \begin{lrbox}{\dratlfscbox}
    \begin{tabular}{c}
\begin{lstlisting}
(program is_specified_drat_proof ((f cnf) (proof DRATProof)) bool
   (match proof
       (DRATProofn (cnf_has_bottom f))
       ((DRATProofa c p) (
           match (is_rat f c) (tt (is_specified_drat_proof (cnfc c f) p)) (ff ff)))
       ((DRATProofd c p) (is_specified_drat_proof (cnf_remove_clause c f) p))))
\end{lstlisting}
    \end{tabular}
    \end{lrbox}
  \scalebox{.9}{\usebox{\dratlfscbox}}
\caption{
  Side condition for checking a specified \drat proof.
  The side conditions
  \texttt{cnf\_has\_bottom}, \texttt{is\_rat}, and \texttt{cnf\_remove\_clause}
  are defined in the same signature.
  Type \texttt{cnfc} is a constructor for CNF formulas, and is
  defined in a separate signature.
}
\label{fig:drat-lfsc}
\end{figure}

In this section, we describe the main characteristics of the \lfsc
signatures\footnote{
\url{https://github.com/CVC4/CVC4/blob/master/proofs/signatures/}}
that we have defined for checking \drat, \lrat, and \er proofs.

The \emph{\lfsc \drat signature} makes extensive use of side conditions to express
processes such as unit propagation and the search for the resolvents of a proper
RAT.
Because of the divergence between operational and specified \drat
and the resulting ambiguity (see \cite{rebola2018two} for further details), 
our signature accepts both kinds of proofs.
\Cref{fig:drat-lfsc} shows the main side condition that is used to check a \drat proof.
Though we do not explain the \lfsc syntax in detail here due to lack of space, the
general idea can be easily understood.  Given a proof candidate \texttt{proof}, it covers three cases:
$(i)$~the proof is empty and the working formula includes a contradiction;
$(ii)$~the proof begins with an addition of a (proper or improper) RAT; or
$(iii)$~the proof begins with a deletion of some clause.
In $(ii)$ and $(iii)$, the same side condition is recursively called on
the rest of \texttt{proof}, with an updated working formula.
In $(ii)$, side condition \texttt{is\_rat} checks whether the added clause is
indeed a RAT via resolvent search and unit propagation.

The \emph{\lfsc \lrat signature} is similar in nature, and 
also makes extensive use of side conditions---albeit
less computationally expensive ones.
In particular, this signature uses hints provided in the \lrat proofs to accelerate
unit propagation.

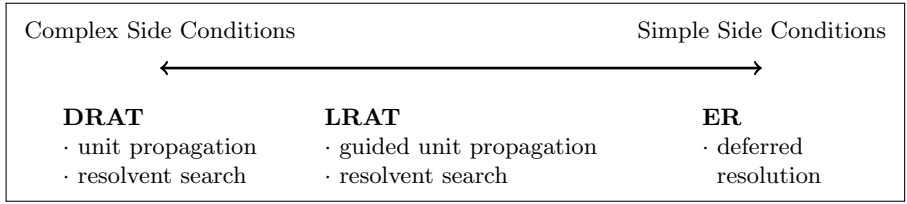
\begin{figure}[t]
  \centering
  \begin{tikzpicture}[framed]
    \draw[line width=1pt,<->]
    (-6,0) node [above=6pt] {Complex Side Conditions}
    -- (2,0) node [above=6pt] {Simple Side Conditions};
    \draw (-6,0) node [below=12pt,align=left] {
      \textbf{\drat}\\
      $\cdot$ unit propagation\\
      $\cdot$ resolvent search
    };
    \draw (-2,0) node [below=12pt,align=left] {
      \textbf{\lrat}\\
      $\cdot$ guided unit propagation\\
      $\cdot$ resolvent search
    };
    \draw (2,0) node [below=12pt,align=left] {
      \textbf{\er}\\
      $\cdot$ deferred\\\hspace{0.6em}resolution
    };
  \end{tikzpicture}
  \vspace*{-2ex}
  \caption{Side conditions across our signatures.}
  \label{fig:sc-complexity}
  \vspace*{-2ex}
\end{figure}

The \emph{\lfsc \er signature} is an extension of 
the \lfsc signature for resolution
proofs that is currently employed by \cvcfour.
It
implements \textit{deferred resolution} to quickly
check large resolution proofs using only a single
side condition~\cite{Stump2013}.
The signature extends resolution in order to check the ER proofs produced by the
\texttt{\drattoer} tool.
These proofs feature extensions of the form
$ x \liff (p \lor (l_1 \land l_2 \land \cdots \land l_k))$, where $x$ is
fresh and $p$ and $l_i$ are not.
Our signature includes side-condition-free rules for introducing such extensions
and translating them to
CNFs of the form
$\set{\set{x,\overline{p}}, \set{x,\overline{l_{1}}\til \overline{l_{k}}},
\set{\overline{x},p,l_{1}}\til \set{\overline{x},p,l_{k}} }$.
The \texttt{decl\_definition} rule in Figure \ref{fig:er-lfsc} is used to
introduce these extensions.
Its first two arguments are literal $p$ and the list of literals
$l_i$ (denoted as \texttt{ls} of type \texttt{lit\_list}) from the
definition. 
The third
argument is a function that receives a fresh variable \texttt{x}
and connects the introduced definition to the rest of the proof.
%
\Cref{fig:sc-complexity} illustrates the difference in side conditions between
the three signatures.

\begin{figure}[t]
  \centering
  \begin{lrbox}{\erlfscbox}
    \begin{tabular}{c}
\begin{lstlisting}
(declare definition (! x var (! p lit (! ls lit_list type))))
(declare decl_definition
   (! p lit (! ls lit_list (! pf_continuation
           (! x var (! def (definition x p ls) (holds empty_clause)))
   (holds empty_clause)))))
\end{lstlisting}
    \end{tabular}
  \end{lrbox}
  \scalebox{0.9}{\usebox{\erlfscbox}}
\caption{Derivation rules for checking that a clause constitutes an extension.}
\label{fig:er-lfsc}
\end{figure}

\section{Workflow: From CVC4 to \lfsc}

\begin{figure}[t]
    \centering
    \begin{mdframed}
      \resizebox{\textwidth}{!}{{
  \tikzstyle{cloud} =
    [draw, ellipse, text width=4em, text centered, node distance=0pt,
     minimum height=3em, inner sep=2pt, fill=yellow!20]
  \tikzstyle{node} =
    [draw, rectangle, rounded corners, text width=6em, text centered,
    inner sep=0.5em, minimum height=3em]
  \tikzstyle{fsmnode} =
    [draw, rectangle, rounded corners, text width=7em, text centered,
     inner sep=0.5em, fill=cyan!20, minimum height=3em]
  \tikzstyle{box} =
    [draw, rectangle, text width=9em, text centered, inner sep=0.5em,
     minimum height=3em, fill=white]

  \begin{tikzpicture}
    \node[text width=6em,text centered] (bvp) {Bit-Vector Formula};
    \node[box,below=10ex of bvp,text width=6em] (bvs) {Bit-Vector Solver};
    \node[node,fill=cyan!20,below=10ex of bvs] (sat) {SAT Solver};

    \node[box,right=10ex of bvs,text width=6em] (bvps) {Bit-Vector Proof Storage};

    \node[node,fill=cyan!20,right=12ex of bvps] (dtrim) {\drattrim};

    \node[node,fill=cyan!20,below=10ex of dtrim] (drater) {\drattoer};

    \node[box,right=16ex of dtrim,text width=6em] (dratp) {\drat\\\lfsc Printer};
    \node[box,above=10ex of dratp,text width=6em] (lratp) {\lrat\\\lfsc Printer};
    \node[box,below=10ex of dratp,text width=6em] (erp) {\er\\\lfsc Printer};

    \node[node,fill=cyan!20,right=6ex of dratp] (lfscc) {\lfsc Checker};

    \node[above=10ex of lfscc] (sig) {\lfsc Signatures};

    \node[above right=-3ex and 4ex of lfscc,text=blue] (ok) {\checkmark};
    \node[below right=-3ex and 4ex of lfscc,text=red] (fail) {\ding{53}};

    \draw[line] (bvp) -- (bvs);
    \draw[line] (bvs) -- node [rotate=90,yshift=1.5ex] {CNF} (sat);
    \draw[line] (bvs) -- node [yshift=1.5ex] {CNF} (bvps);
    \draw[line] (sat) -| node [xshift=-12ex,yshift=1.5ex] {\drat} (bvps);

    \draw[line] (bvps) -- node [xshift=1.5ex,text width=4em] {CNF+ \drat} (dtrim);

    \draw[line] (dtrim) |- node [xshift=16.5ex,yshift=1.5ex] {CNF+\lrat} (lratp);
    \draw[line] (dtrim) -- node [yshift=1.5ex] {CNF+\drat} (dratp);
    \draw[line] (dtrim) -- node [rotate=90,xshift=1.5ex,text width=4em] {CNF+ \drat} (drater);
    \draw[line] (drater) -- node [yshift=1.5ex] {CNF+\er} (erp);

    \draw[line] (lratp.0) -| +(1em,0ex) |- (lfscc);
    \draw[line] (dratp) -- (lfscc);
    \draw[line] (erp.0) -| +(1em,0ex) |- (lfscc);

    \draw[line] (sig) -- (lfscc);
    \draw[line] (lfscc.0) +(0ex,2.25ex) -- (ok);
    \draw[line] (lfscc.0) +(0ex,-2.25ex)-- (fail);
  \end{tikzpicture}
}}
    \end{mdframed}
    \vspace*{-2ex}
    \caption{
      Producing and checking \lfsc proofs in \drat, \lrat and \er proof systems
      in \cvcfour.
      White boxes are \cvcfour-internal components; blue boxes are
      external libraries.
    }
    \label{fig:workflow}
    \vspace*{-2ex}
  \end{figure}
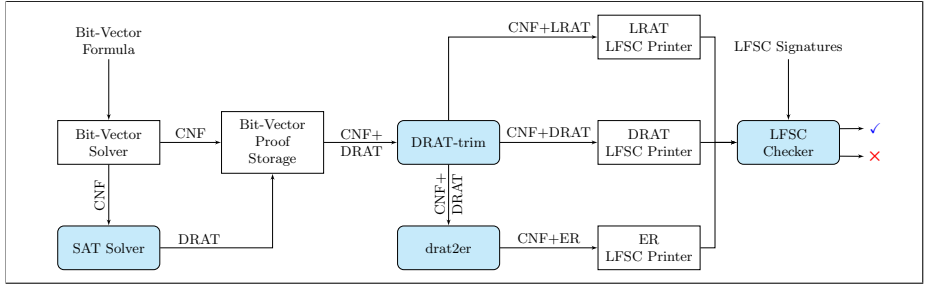

\Cref{fig:workflow} shows the general workflow for incorporating \drat proofs
in the \lfsc proof infrastructure of \cvcfour after bit-blasting.
\lfsc proofs for the bit-blasting step are described in
\cite{DBLP:conf/lpar/HadareanBRTD15}.
A given bit-vector formula
is bit-blasted to SAT, and the resulting CNF is then sent to the underlying
SAT solver.
We use \drattrim to
trim the original formula,
optimize the proof produced by the SAT solver,
and optionally produce an \lrat proof that is
forwarded to the \lrat \lfsc pipeline.
In case of \drat \lfsc proofs, we can directly use the optimized proof and
formula emitted by \drattrim.
For \er \lfsc proofs, we first use \drattoer to translate the optimized \drat
proof into an \er proof, which is then sent to the \er \lfsc pipeline.
The result of each pipeline is an \lfsc proof in the corresponding proof
system, which can be checked with the corresponding signature
(see \Cref{sec:sigs}) using the \lfsc proof checker.
Note that prior to bit-blasting,
the input is usually simplified via rewriting and other preprocessing
techniques, 
for which \cvcfour currently does not produce proofs.
The addition of such proofs is left as future work and orthogonal to
incorporating
\drat proofs from the SAT solver back-end,
which is the focus of this paper.


\section{Experiments}

\begin{table}[b]
  \centering
  \scalebox{.9}{{%
  \renewcommand{\arraystretch}{1.2}%
  \setlength{\tabcolsep}{5pt}
  \begin{tabular}{|l|rr|rr|rr|rr|}
    \hline
    \textbf{Proof System}
    & \multicolumn{2}{c|}{\textbf{solve}}
    & \multicolumn{2}{c|}{\textbf{log}}
    & \multicolumn{2}{c|}{\textbf{log,prod}}
    & \multicolumn{2}{c|}{\textbf{log,prod,check}}
    \\
    & \# & [s]
    & \# & [s]
    & \# & [s]
    & \# & [s]
    \\
    \hline
    \hline
    Resolution & 20480                  & 464k                  & 20396                  & 524k                  & \textbf{14217} & \textbf{4400k} & 13510 & 4973k \\
    \hline
    \drat      & \multirow{3}{*}{\textbf{20767}} & \multirow{3}{*}{\textbf{283k}} & \multirow{3}{*}{\textbf{20736}} & \multirow{3}{*}{\textbf{319k}} & 14098 & 4492k & 12563 & 5616k \\
    \cline{6-9}
    \lrat      &                        &                       &                        &                       & 14088 & 4500k & 12877 & 5370k \\
    \cline{6-9}
    \er        &                        &                       &                        &                       & 14035 & 4565k & \textbf{13782} & \textbf{4886k} \\
    \hline
  \end{tabular}
}
}
  \vspace{1ex}
  \caption{Impact of proof logging, production, and checking on \# solved
  problems.}
  \label{tab:evaluation}
\end{table}

We implemented the three approaches described in \Cref{sec:integration} in
\cvcfour using \cryptominisat 5.6~\cite{cryptominisat} as the SAT back-end.
We compared them against the resolution-based proof machinery currently
employed in \cvcfour 
and evaluated our techniques on all 21125 benchmarks from the quantifier-free
bit-vector logic QF\_BV of SMT-LIB~\cite{SMTLib2010} with status
\emph{unsat} or \emph{unknown}.
%
%
%
All experiments were performed on a cluster with Intel Xeon E5-2620v4 CPUs
with 2.1GHz and 128GB of memory.
We used a time limit of 600 seconds (CPU time)
and a memory limit of 32GB for each solver/benchmark pair.
For each error or memory-out, we added a penalty of 600 seconds.
%

\begin{wrapfigure}{r}{0.45\textwidth}
  \vspace{-5ex}
  \centering
  \includegraphics[width=0.44\textwidth]{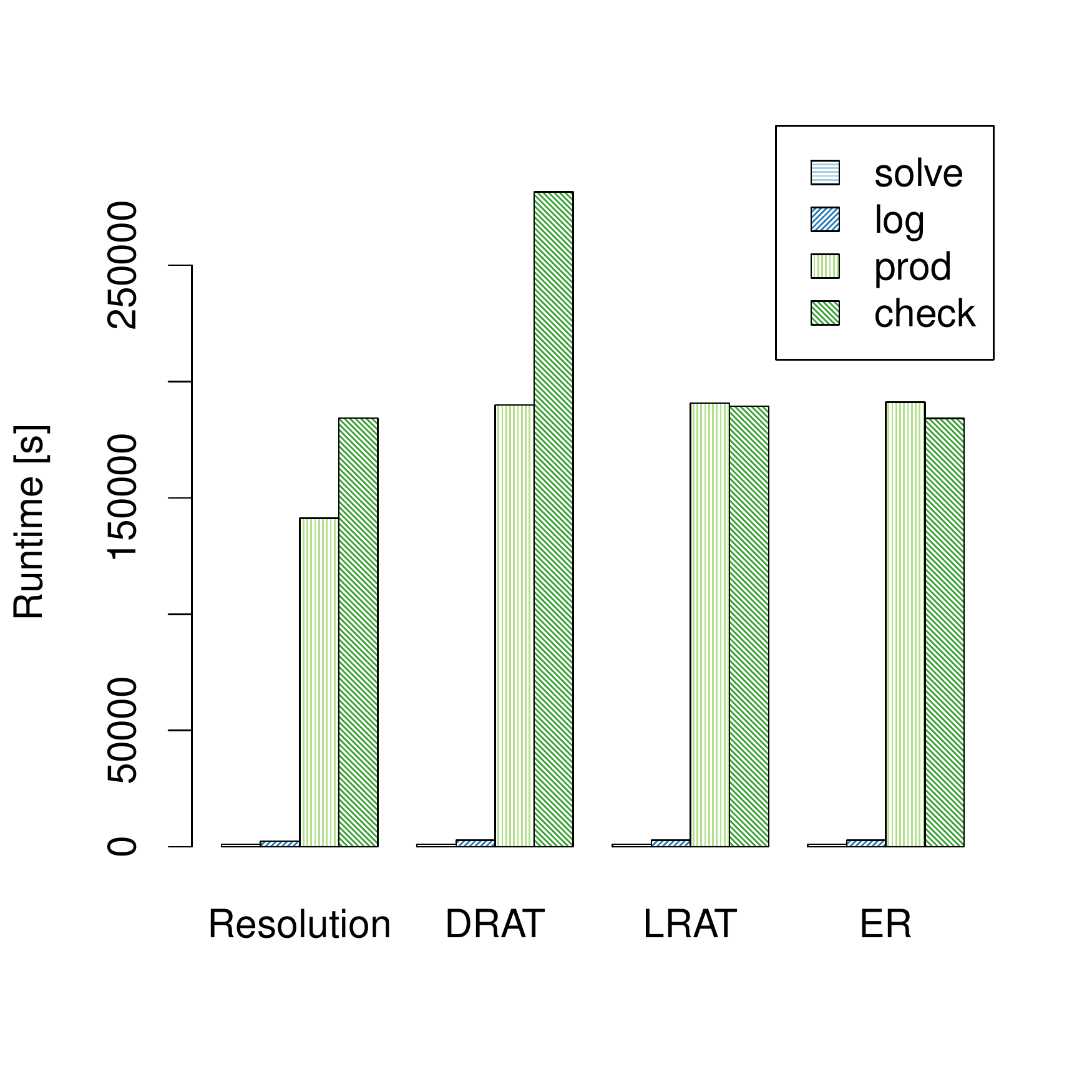}
  \vspace{-6ex}
  \caption{
    Runtime distribution on 12539 commonly proved problems.
  }
  \label{fig:plot}
  \vspace{-4ex}
\end{wrapfigure}

\Cref{tab:evaluation} shows the results for the Resolution approach with
\minisat,
and the \drat, \lrat and \er approaches with \cryptominisat.
For each, we ran the following four configurations:
proofs disabled (\emph{solve}),
proof logging enabled (\emph{log}),
proof production enabled (\emph{prod}),
and proof checking enabled (\emph{check}).
%
Proof logging records proof-related
information but does not produce the actual proof,
e.g., when producing \drat proofs, proof logging stores
the \drat proof from the SAT-solver,
which is only translated to \lfsc during proof production.
%
In the \emph{solve} configuration, the \drat-based approaches (using \cryptominisat)
solve 287 more problems than the Resolution approach (which uses \cvcfour's
custom version of MiniSat).
This indicates that the custom version of MiniSat was a bottleneck for solving.
In the \emph{log} configuration, the \drat-based approaches 
solve 31 fewer problems than in the \emph{solve} configuration;
and in
the \emph{prod} configuration 
the \drat-based approaches produce proofs for $\sim$6600
fewer problems.
This indicates that the bottleneck in the \drat-based approaches is 
the translation of \drat to
\lfsc.
For all approaches, about 30\% of the solved problems require more than 8GB of
memory to produce a proof, showing that proof production can in
general be very memory-intensive.
Finally, with proof checking enabled, the \er-based approach outperforms all other
approaches.
Note that in $\sim$270 cases,
\cryptominisat produced a \drat proof that was
rejected by \drattrim, which we counted as error.
Further, for each \emph{check} configuration, our \lfsc checker reported $\sim$200 errors,
which are not related to our new approach.
Both issues need further investigation.

\Cref{fig:plot} shows the runtime distribution 
for all approaches and configurations over the commonly proved problems
(12539 in total).
The runtime overhead of proof production for the \drat-based approaches is
1.35 times higher compared to resolution.
This is due to the fact that we post-process the \drat-proof
prior to translating it to \lfsc, which involves
writing temporary files and calling external libraries.
The proof checking time correlates with the complexity of the side conditions
(see \Cref{fig:sc-complexity}),
where \er clearly outperforms \drat.
%

%

\section{Conclusion}

We have described three approaches for integrating \drat proofs in \lfsc, 
which enable us to use off-the-shelf SAT solvers as the
SAT back-end for the bit-blasting engine of \cvcfour while supporting bit-vector
proofs.
%
%
For future work,
we plan to reduce the complexity of the side conditions in the \drat and \lrat
signatures
%
and
the proof
production overhead in the translation workflows.
%
We also
plan to add support for the new signatures in \smtcoq \cite{DBLP:conf/cav/EkiciMTKKRB17},
a tool that increases automation in \coq \cite{coqref} using proofs generated
by \cvcfour.
In a more applicative direction, we plan to explore the potential 
\drat proofs in SMT-solvers may have in the proof-carrying code paradigm
\cite{DBLP:conf/popl/Necula97},
as well as its recent variant in blockchains, namely proof-carrying
smart contracts \cite{DBLP:conf/fc/DickersonGHSK18}.

%
%
%
 \bibliographystyle{splncs04}
 \bibliography{mybib}
\end{document}